\documentclass[journal,twoside,web]{ieeecolor}
\usepackage{generic}
\usepackage{amsmath,amssymb,amsfonts}
\usepackage{algorithmic}
\usepackage{graphicx}
\usepackage{algorithm,algorithmic}
\usepackage{textcomp}

\usepackage{booktabs}
\usepackage{siunitx}
\usepackage{float}
\usepackage{array}
\usepackage{subcaption}
\usepackage{bm}
\usepackage[hidelinks]{hyperref}
\usepackage[
    backend=biber,
    style=ieee,
    sorting=none,
    doi=false,
    url=false,
    isbn=false,
    maxnames=6, 
    minnames=1
]{biblatex}
\def\BibTeX{{\rm B\kern-.05em{\sc i\kern-.025em b}\kern-.08em
    T\kern-.1667em\lower.7ex\hbox{E}\kern-.125emX}}
\usepackage{fancyhdr}
\emergencystretch=\maxdimen
\begin{document}
\title{\textbf{Explainable Deep Learning of Resting-State Functional Connectomes Reveals Network Biomarkers of Adolescent Intelligence}}
\author{
Md. Tanvir Rahman,
Nabil Anan Orka,
Asaduzzaman Khan, and
Mohammad Ali Moni
\thanks{Md. Tanvir Rahman, Nabil Anan Orka, Asaduzzaman Khan, and Mohammad Ali Moni are with the School of Health and Rehabilitation Sciences, The University of Queensland, QLD 4072, Australia.}
\thanks{Md. Tanvir Rahman is also with the Department of Information and Communication Technology, Mawlana Bhashani Science and Technology University, Tangail 1902, Bangladesh.}
}

\maketitle
\thispagestyle{fancy}

\begin{abstract}
Mapping resting-state brain organization to individual differences in cognitive ability remains a major challenge in population neuroinformatics. Although deep learning enables flexible modeling of brain connectivity, limited interpretability restricts its scientific and clinical utility. To address this objective, we developed an explainable deep learning framework based on sparse projected residual networks to predict fluid, crystallized, and total intelligence from resting-state functional magnetic resonance imaging in 5,285 participants from the Adolescent Brain Cognitive Development study. We incorporated three complementary explainability methods (Integrated Gradients, Gradient Shapley Additive Explanations, and Occlusion) to interpret model behavior. The framework outperformed existing approaches, achieving Pearson correlations of 0.44, 0.58, and 0.56 for fluid, crystallized, and total intelligence, respectively, corresponding to predictive improvements of 6 to 9 percent. All three explainability methods produced near-identical feature rankings (pairwise rank correlations greater than 0.99). Consensus maps revealed a dual-layered functional architecture where primary predictive hubs localized within canonical systems, while the strongest global predictive pathways frequently bypassed these hubs through distributed, long-range relay connections. These findings suggest that intelligence emerges from the interaction between localized computational hubs and distributed communication pathways. Ultimately, these normative network architectures provide clinical reference maps to detect individual deviations, supporting earlier diagnosis, cognitive subtype stratification, and treatment monitoring in atypical neurodevelopment.
\end{abstract}


\section{Introduction}
\label{sec:introduction}

\IEEEPARstart{B}{rain} mapping aims to explain how cognitive abilities emerge from distributed patterns of neural activity \cite{bassett2017network}. Rather than operating within isolated regions, higher-order processing relies on coordinated interactions across large-scale functional networks within human physiological systems \cite{Santoro2024}, making the characterization of these network-level phenotypes essential for understanding individual neurocognitive differences. Resting-state functional magnetic resonance imaging (rs-fMRI) provides a scalable framework for this goal by measuring spontaneous whole-brain activity. Functional connectomes derived from rs-fMRI capture stable population-level organization while preserving subject-specific variability, and have been shown to predict cognitive traits \cite{Fu2023, Omidvarnia2024} and disorders \cite{Noman2024}.

Deep learning (DL) offers a powerful approach for modeling such high-dimensional data by learning hierarchical representations directly from connectomes \cite{ravindran2022five, Li_2024, Zhang_2024}. Recent DL-based methods have shown promising performance in predicting cognitive scores from rs-fMRI data \cite{my_systematic_review}. Broadly, these approaches fall into two categories. The first leverages connectivity-based representations, using resting-state functional connectivity matrices as inputs and employing convolutional or graph-based architectures to model interactions between brain regions \cite{fc_model3, Xia_2023, Huang_2022}. The second focuses on spatio-temporal modeling, operating directly on resting-state time series to capture dynamic brain activity patterns, typically via convolutional and recurrent networks \cite{st_model1, st_model2, Li_2023}.

Despite substantial progress, reliable neurocognitive prediction from rs-fMRI remains challenging. The representation of the input signal heavily dictates predictive performance, yet contemporary deep learning literature frequently overemphasizes architectural permutations at the expense of rigorous input engineering. Consequently, high-dimensional connectomic features relative to sample cohorts amplify overfitting risks and introduce substantial information leakage vectors \cite{cearns2019recommendations, cui2018effect, ambroise2002selection}. Moreover, demographic and developmental factors contribute to structured variance that should be explicitly modeled \cite{rao2017predictive}. Within this context, the limited interpretability of DL models, due to their black-box nature, remains a critical barrier to scientific and clinical adoption. In neuroimaging, interpretability helps verify that model predictions reflect biologically meaningful signals rather than spurious correlations. Although explainable artificial intelligence (XAI) has been proposed to address this challenge, its use in neuroimaging remains relatively limited and often relies on a single attribution method, which makes explanations sensitive to method-specific biases \cite{atrribution_sensitivity}.

In this work, we present an explainable DL framework for neurocognitive prediction from resting-state functional connectomes. The pipeline integrates strictly nested, bootstrapped consensus feature selection with a Sparse Projected Residual Network (SPRN) adapted for high-dimensional neuroimaging data. Demographic covariates are incorporated via late fusion to avoid leakage, and evaluation follows recommended best practices for large-scale neuroimaging studies \cite{abcd_guideline}. Crucially, we employ an explainability convergence framework to enhance the reliability of model interpretation. Instead of relying on a single XAI method, we employ three methodologically distinct approaches spanning gradient-based attribution and perturbation analysis (Integrated Gradients \cite{integrated_gradients}, GradientSHAP \cite{shap}, and Occlusion \cite{occlusion}) and quantify their agreement using rank-based metrics. Beyond validating consistency, feature importance scores systematically quantify the contributions of functional connectivity edges and regional variance features, which are then aggregated into node- and network-level summaries. This enables the identification of stable, biologically meaningful interaction patterns, addressing a key limitation of prior XAI studies in neurocognition.

\begin{table*}[t]
\caption{Demographic and Cognitive Score Summary Across Five Family-Aware Cross-Validation Folds. Mean \(\pm\) SD are reported.}
\label{tab:demographics}
\centering
\begin{tabular*}{\textwidth}{@{\extracolsep{\fill}} l l l l l l l l l}
\toprule
\textbf{Fold} & \textbf{Split} & \textbf{Subjects} & \textbf{Males} & \textbf{Females} & \textbf{Age (months)} & \textbf{\bm{$G_f$}} & \textbf{\bm{$G_c$}} & \textbf{\bm{$G_t$}} \\
\midrule
1 & Train & 3758 & 1821 & 1936 & 119.86 $\pm$ 7.50 & 93.02 $\pm$ 10.09 & 87.23 $\pm$ 6.67 & 87.60 $\pm$ 8.52 \\
 & Validation & 470 & 239 & 231 & 119.77 $\pm$ 7.46 & 93.69 $\pm$ 9.64 & 87.49 $\pm$ 6.64 & 88.16 $\pm$ 8.20 \\
 & Test & 1057 & 500 & 557 & 119.66 $\pm$ 7.66 & 93.01 $\pm$ 10.06 & 87.12 $\pm$ 6.56 & 87.54 $\pm$ 8.44 \\
\midrule
2 & Train & 3765 & 1814 & 1950 & 119.73 $\pm$ 7.53 & 92.96 $\pm$ 10.13 & 87.18 $\pm$ 6.66 & 87.53 $\pm$ 8.56 \\
 & Validation & 463 & 250 & 213 & 120.06 $\pm$ 7.48 & 92.67 $\pm$ 10.18 & 87.68 $\pm$ 6.86 & 87.65 $\pm$ 8.58 \\
 & Test & 1057 & 496 & 561 & 119.98 $\pm$ 7.53 & 93.65 $\pm$ 9.67 & 87.21 $\pm$ 6.48 & 88.00 $\pm$ 8.15 \\
\midrule
3 & Train & 3758 & 1816 & 1941 & 119.85 $\pm$ 7.53 & 93.09 $\pm$ 9.95 & 87.13 $\pm$ 6.68 & 87.59 $\pm$ 8.43 \\
 & Validation & 470 & 233 & 237 & 119.49 $\pm$ 7.54 & 92.80 $\pm$ 10.36 & 87.42 $\pm$ 6.36 & 87.59 $\pm$ 8.55 \\
 & Test & 1057 & 511 & 546 & 119.81 $\pm$ 7.50 & 93.15 $\pm$ 10.25 & 87.50 $\pm$ 6.64 & 87.82 $\pm$ 8.61 \\
\midrule
4 & Train & 3762 & 1813 & 1949 & 119.86 $\pm$ 7.58 & 93.26 $\pm$ 10.05 & 87.26 $\pm$ 6.63 & 87.77 $\pm$ 8.47 \\
 & Validation & 466 & 228 & 238 & 119.27 $\pm$ 7.39 & 92.95 $\pm$ 10.05 & 87.13 $\pm$ 6.36 & 87.48 $\pm$ 8.28 \\
 & Test & 1057 & 519 & 537 & 119.88 $\pm$ 7.37 & 92.48 $\pm$ 10.02 & 87.14 $\pm$ 6.82 & 87.23 $\pm$ 8.58 \\
\midrule
5 & Train & 3755 & 1816 & 1938 & 119.86 $\pm$ 7.53 & 93.05 $\pm$ 10.01 & 87.20 $\pm$ 6.62 & 87.61 $\pm$ 8.44 \\
 & Validation & 473 & 210 & 263 & 119.63 $\pm$ 7.38 & 93.26 $\pm$ 10.00 & 87.58 $\pm$ 6.69 & 87.96 $\pm$ 8.50 \\
 & Test & 1057 & 534 & 523 & 119.73 $\pm$ 7.58 & 93.09 $\pm$ 10.22 & 87.16 $\pm$ 6.71 & 87.60 $\pm$ 8.60 \\
\bottomrule
\end{tabular*}
\vspace{2mm}

\parbox{1\textwidth}{\small \textit{Note:} One participant identified as intersex was retained via one-hot encoding.}
\end{table*}

\section{Methods}
\label{sec:methods}
\begin{figure*} [!t]
	\centering
	\includegraphics[width=\textwidth]{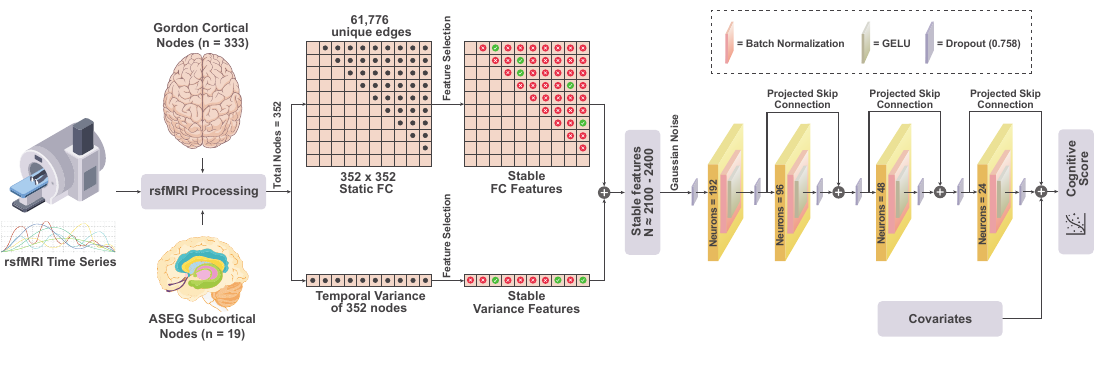}
    \caption{Overview of the proposed methodology. Resting-state networks are constructed from 352 ROIs (333 cortical, 19 subcortical), yielding 61,776 static Functional Connectivity edges and 352 temporal variance features. Following family-aware stratification, bootstrapped Ridge consensus masking (50\% threshold) reduces the feature space to approximately 2,100--2,400 stable features. The SPRN predicts cognitive scores through learned residual projections across descending latent dimensions (192, 96, 48, 24) with Gaussian noise augmentation, multi-level dropout ($p = 0.758$), Batch Normalization, and GELU activations. Covariates are incorporated via late fusion to prevent confounding of neural representations.}
	\label{fig:methodology}
\end{figure*}

We developed a multi-stage pipeline for leakage-free prediction of cognitive ability from resting-state functional connectivity. An overview is provided in Fig. \ref{fig:methodology}.

\subsection{Dataset and Cohort Selection}
We used data from the Adolescent Brain Cognitive Development (ABCD) study \cite{abcd}, the largest long-term study of brain development and child health in the United States. The cohort includes over 11,000 children aged 9-10 years at baseline, recruited across 21 sites with standardized multimodal MRI on three scanner platforms (Siemens Prisma, GE 750, Philips Achieva). Below, we detail the cognitive targets, confound control strategy, and inclusion criteria.

\subsubsection{Cognitive Phenotyping (Target Variables)}
Cognitive targets were derived from the NIH Toolbox Cognition Battery \cite{abcd}. We modeled three intelligence domains: crystallized intelligence (\(G_c\)), reflecting language and acculturated knowledge; fluid intelligence (\(G_f\)), representing novel problem-solving, executive function, working memory, and processing speed; and total intelligence (\(G_t\)), a global composite measure of general cognitive ability (\(g\)). Uncorrected standard scores were used to preserve developmental variability across participants. To improve optimization stability, scores were quantile-transformed prior to training, while all predictions were inverse-transformed back to the original scale for evaluation, enabling accurate estimation of \(R^2\) on the true developmental variance.

\subsubsection{Confound Control and Covariates}
To ensure models capture neural substrates rather than demographic or acquisition artifacts, we controlled for nuisance variables following ABCD recommendations \cite{abcd_guideline}: (i) age in months, (ii) sex at birth, (iii) MRI device serial number (scanner-specific differences), (iv) estimated total intracranial volume (global brain size effects), and (v) mean framewise displacement of retained frames (motion-driven artifacts). Categorical variables were one-hot encoded and continuous variables were standardized, yielding a final covariate vector $\mathbf{c}_i \in \mathbb{R}^{24}$.

\subsubsection{Inclusion Criteria}
Subjects were retained based on recommended QC for ABCD rs-fMRI data (Release 5.1) \cite{abcd_guideline}: (i) complete cognitive and covariate data, (ii) a \textit{pass} rating from the ABCD-HCP BIDS QC pipeline, and (iii) at least 10 minutes (approximately 750 frames) of low-motion data after frame censoring (Framewise Displacement $< 0.2$ mm). The final cohort consisted of 5,285 adolescents. Because the ABCD study includes related individuals, family IDs were used to enforce family-wise separation during cross-validation and prevent information leakage \cite{abcd_guideline}. Demographic statistics and cognitive-score distributions across the five folds are summarized in Table \ref{tab:demographics}.

\subsection{fMRI Preprocessing and Feature Engineering}
Resting-state fMRI data were preprocessed using the ABCD-HCP-BIDS pipeline (DCAN Labs, ABCC repository) \cite{ABCC}, including motion correction, distortion correction, spatial normalization to MNI space, and frame censoring (FD~${>}0.2$~mm). For reproducibility, we used standardized derived measures from ABCD Release 5.1. We constructed a hybrid feature space spanning $R = 352$ regions of interest (ROIs): the Gordon \cite{gordon} cortical parcellation ($n = 333$) and ASEG \cite{ASEG} subcortical structures ($n = 19$). Two complementary feature representations were extracted from the preprocessed time series: static functional connectivity and node-wise temporal variance.

\subsubsection{Static Functional Connectivity (sFC)}
Subject-specific dense functional connectivity matrices were obtained from the ABCC repository, representing Fisher's $z$-transformed Pearson correlations between all ROI pairs computed on the frame-censored time series. For each subject $i$, upper triangular elements of the $352 \times 352$ matrix $\mathbf{A}^{(i)}$ was extracted to form $\mathbf{x}_{conn} \in \mathbb{R}^{E}$, where $E = 61,776$ unique edges.

\subsubsection{Node Signal Variance}
Pre-tabulated temporal variance ($\sigma^2$) of the blood oxygen level dependent (BOLD) signal for each ROI was obtained from the ABCD Release 5.1, capturing local fluctuation amplitude independent of network coupling. These were aggregated into a node-wise feature vector $\mathbf{x}_{var} \in \mathbb{R}^{R}$.

The final input vector for subject \(i\) is the concatenation \(\mathbf{x}_{raw}^{(i)} = [\mathbf{x}_{conn} \oplus \mathbf{x}_{var}]\), resulting in a total of \(62,\!128\) features per subject (\(61,\!776\) edges and \(352\) node-wise variance features).

\subsection{Robust Bootstrapped Feature Selection}
Given the high dimensionality of the feature space relative to sample size, we implemented a nested, group-aware bootstrapped selection framework applied strictly within the training indices of each cross-validation fold to prevent data leakage.

\subsubsection{Group-Aware Bootstrapped Ridge Regression}
Within each training fold, we identified the optimal $L_2$ penalty ($\alpha$) via internal Group $K$-Fold cross-validation over a grid $\alpha \in [100, 50000]$, preserving family structures. Using this $\alpha$, we performed $B = 100$ bootstrap iterations, sampling families with replacement. A Ridge model was fit in each iteration, and features were deemed active if their absolute coefficient exceeded a dynamic threshold:
\begin{equation}
    \tau_b = \mu(|\boldsymbol{\beta}_b|) + 1.5 \cdot \sigma(|\boldsymbol{\beta}_b|)
\end{equation}
where $\mu$ and $\sigma$ are the mean and standard deviation of absolute coefficients for bootstrap $b$. The $1.5\sigma$ multiplier balances aggressive sparsification against retention of consistently predictive coefficients. $B = 100$ provided stable empirical estimation while remaining computationally tractable within the nested cross-validation framework.

\begin{table}[h]
\caption{Average Number of Stable Features Retained Across 5 Folds by Consensus Threshold. Mean $\pm$ SD are reported.}
\label{tab:feature_counts}
\centering
\renewcommand{\arraystretch}{1.2}
\begin{tabular*}{\columnwidth}{@{\extracolsep{\fill}} l l l l}
\toprule
\textbf{Threshold} & \textbf{\bm{$G_f$}} & \textbf{\bm{$G_c$}} & \textbf{\bm{$G_t$}} \\
\midrule
30\% & 5015.2 $\pm$ 54.3 & 5281.0 $\pm$ 72.8 & 5200.6 $\pm$ 43.1 \\
40\% & 3230.0 $\pm$ 54.9 & 3564.4 $\pm$ 49.3 & 3490.2 $\pm$ 42.6 \\
50\% & 2127.0 $\pm$ 20.9 & 2394.8 $\pm$ 10.3 & 2361.4 $\pm$ 53.7 \\
60\% & 1366.6 $\pm$ 15.6 & 1610.6 $\pm$ 20.0 & 1585.6 $\pm$ 53.3 \\
70\% & 855.0 $\pm$ 11.5 & 1023.2 $\pm$ 26.6 & 1013.8 $\pm$ 34.6 \\
80\% & 468.2 $\pm$ 19.9 & 599.0 $\pm$ 21.6 & 582.2 $\pm$ 25.5 \\
\bottomrule
\end{tabular*}

\end{table}

\subsubsection{Consensus Masking}
A feature was retained only if active in $\geq 50\%$ of bootstraps, producing a binary selection mask $\mathbf{M}$ for that fold, applied to both $\mathbf{x}_{\text{conn}}$ and $\mathbf{x}_{\text{var}}$. This mask served as a fixed stencil for validation and test sets within the same fold. As detailed in Table \ref{tab:feature_counts}, this threshold effectively filtered out high-dimensional topological noise, isolating a highly stable subset comprising approximately 3.5\% of the original 62,128 multimodal features across all target domains. This distinct fold-to-fold consistency demonstrates that the masking pipeline isolates robust neurophysiological signatures rather than sample-specific noise, ensuring the clinical generalizability of the downstream predictive biomarkers.

\subsection{Sparse Projected Residual Network (SPRN)}
We introduce SPRN, designed for highly pruned tabular neuroimaging representations. SPRN employs a funnel-style residual topology with aggressive regularization to combat representation degradation and overfitting.

\subsubsection{Architecture and Regularization}
The network receives the masked input through dropout ($p = 0.758$) and projects it to a 192-neuron latent space, then descends through $L = 4$ residual blocks with halving dimensionality ($192 \rightarrow 96 \rightarrow 48 \rightarrow 24$). Continuous data augmentation via Gaussian noise injection ($\mathcal{N}(0, 0.2463^2)$) was applied element-wise at every training epoch. Within each block, representations undergo Batch Normalization, GELU activation, and dropout ($p = 0.758$).

\subsubsection{The Projected Residual Mechanism}
SPRN replaces standard identity mappings with learned linear projections, enabling skip connections across decreasing dimensions. For a transition from block $l$ ($\mathbf{h}_l \in \mathbb{R}^{d_{\text{in}}}$) to block $l+1$ ($\mathbf{h}_{l+1} \in \mathbb{R}^{d_{\text{out}}}$):
\begin{equation}
    \tilde{\mathbf{h}}_{l+1} = \text{Dropout}(\text{GELU}(\text{BN}(\mathbf{W}_l \mathbf{h}_l + \mathbf{b}_l)))
\end{equation}
\begin{equation}
    \mathbf{h}_{l+1} = \tilde{\mathbf{h}}_{l+1} + (\mathbf{W}_{\text{proj}, l} \mathbf{h}_l + \mathbf{b}_{\text{proj}, l})
\end{equation}
where $\mathbf{W}_{\text{proj}, l} \in \mathbb{R}^{d_{\text{out}} \times d_{\text{in}}}$ matches dimensions while maintaining the residual path, forcing the feed-forward path to learn strict residual refinements.

\subsubsection{Late Fusion of Covariates}
To prevent demographic and imaging covariates from dominating early neural representations, the standardized covariate vector ($\mathbf{c}_i \in \mathbb{R}^{24}$) bypasses all residual blocks and is concatenated only at the terminal readout:
\begin{equation}
    \label{eq:late_fusion}
    \hat{y}_i = \mathbf{W}_{\text{head}} (\mathbf{h}_{\text{final}} \oplus \mathbf{c}_i) + b_{\text{head}}
\end{equation}
where $\mathbf{h}_{\text{final}} \in \mathbb{R}^{24}$ and $\oplus$ denotes concatenation.

\subsection{Training and Evaluation Protocol}
Models were evaluated under strict Group $K$-Fold ($K = 5$) cross-validation using family ID to prevent genetic data leakage between folds. Uncorrected cognitive scores were quantile-transformed to a standard normal distribution before training. Predictions were inverse-transformed for evaluation, enabling accurate calculation of Mean Squared Error (MSE), Mean Absolute Error (MAE), Pearson's $r$, and the Coefficient of Determination ($R^2$).

Hyperparameters (learning rate, weight decay, dropout probability, Gaussian noise standard deviation) were determined via Bayesian optimization (Optuna \cite{optuna}, 100 trials), maximizing mean validation $R^2$. Training minimized MSE loss using AdamW (learning rate $= 8.7 \times 10^{-4}$, weight decay $= 7.1 \times 10^{-3}$). A ReduceLROnPlateau scheduler halved the learning rate if validation $R^2$ stagnated. Early stopping was enforced with a patience of 30 epochs based on validation performance, with a maximum of 300 epochs.

\subsection{Ablation Framework and Hypothesis Testing}
We evaluated eleven distinct ablations and baselines across four categories to decompose SPRN's predictive power.

\subsubsection{Feature Selection Sensitivity}
To assess the 50\% consensus threshold, we trained SPRN across mask stringencies, retaining features active in 30\%, 40\%, 60\%, 70\%, and 80\% of bootstraps. An upper-bound ablation used 100\% of features (no selection) to quantify the cost of unpruned high-dimensional connectomes.

\subsubsection{Architectural Ablations}
\paragraph{SPRN without Residual Connection}
Removed projection matrices ($\mathbf{W}_{\text{proj}}$), reducing the model to a standard deep MLP to test the necessity of skip connections.

\paragraph{SPRN without Variance Data}
Excluded $\mathbf{x}_{\text{var}}$, predicting purely from $\mathbf{x}_{\text{conn}}$ to assess the independent value of local amplitude signals.

\subsubsection{Null-Brain Covariate Isolation}
To confirm SPRN extracts brain-based predictive variance, we executed a ``Covariates Only'' ablation: neuroimaging inputs were dynamically replaced with pure Gaussian noise, forcing the architecture to rely solely on late-fused covariates ($\mathbf{c}_i$).

\subsubsection{Traditional and Graph-Based Benchmarks}
We compared the SPRN against standard regressors using the identical 50\% consensus mask.
Ridge ($\alpha=5000$), Lasso ($\alpha=5000$), ElasticNet ($\alpha=0.01$, $\text{L1\_ratio}=0.15$), Random Forest (100 trees, max depth 10), and histogram-based Gradient Boosting (200 iterations) were evaluated with fixed random seeds.
In addition, we implemented a Graph Convolutional Network (GCN) to test whether spectral message-passing offers advantages over tabular residual modeling.
For the GCN, the masked edge features were first reconstructed into a full $352 \times 352$ adjacency matrix by placing each retained edge into its corresponding upper-triangular position and mirroring to the lower triangle; self-loops were added to ensure numerical stability during message passing. Node features were the 352 temporal variance values. The GCN architecture strictly mirrored the SPRN hyperparameters, and covariates were fused after global mean pooling over nodes.

\subsection{Explainable AI Consensus Framework}
To mitigate individual algorithmic biases and ensure interpretability, we introduced an explainability consensus framework that evaluates feature attributions across three distinct paradigms. Specifically, we computed: (i) \textit{Integrated Gradients (IG)}, which integrates gradients along a linear path from an empirical training-fold baseline to the subject's input; (ii) \textit{GradientSHAP}, which approximates SHAP values via expected gradients over a background reference distribution of 100 random training subjects; and (iii) \textit{Occlusion}, a perturbation-based approach that slides a zero-mask window over individual features to measure the resultant decrement in predictive performance.

Raw attributions were summarized by the absolute mean per feature per method, Min-Max scaled to $[0, 1]$, and averaged across the three explainers to compute a final structural Consensus Score. Methodological convergence was verified across active features for each explainer pair using four complementary metrics: Spearman rank correlation ($\rho$) for monotonic rank agreement, Kendall’s $\tau$ for strict pairwise consistency, Pearson correlation for magnitude agreement, and Jaccard similarity of the top-100 features to quantify critical pathway overlap. These scores were averaged across all five folds to generate global spatial importance maps, remapping edge attributions to their topological source and target indices. Node-level criticality was quantified by summing the consensus attributions of all incident edges for a given region, augmented by the node's intrinsic temporal variance attribution. Crucially, because covariates are incorporated exclusively via terminal late fusion (Eq.~\ref{eq:late_fusion}), gradients computed with respect to the input layer ($\mathbf{x}_{\text{conn}}$ and $\mathbf{x}_{\text{var}}$) remain mathematically isolated from nuisance variables, guaranteeing that the resulting XAI mappings reflect strictly unconfounded neural phenomena.

\begin{table*}[t]
\caption{Ablation Study on Architectural Components of Sparse Projected Residual Networks. Mean $\pm$ SD across 5 Folds is reported. Bold indicates best per target.}
\label{tab:sprn_ablation}
\centering
\renewcommand{\arraystretch}{1.2}
\begin{tabular*}{\textwidth}{@{\extracolsep{\fill}} l l c c c c}
\toprule
\textbf{Target} & \textbf{SPRN Configuration} & \textbf{\bm{$R^2 \uparrow$}} & \textbf{r $\bm{\uparrow}$} & \textbf{MAE $\bm{\downarrow}$} & \textbf{MSE $\bm{\downarrow}$} \\
\midrule
\textbf{\bm{$G_f$}} 
& SPRN (Proposed) & \textbf{0.1906 $\pm$ 0.0175} & \textbf{0.4421 $\pm$ 0.0198} & \textbf{7.1437 $\pm$ 0.0850} & \textbf{81.5946 $\pm$ 1.7941} \\
& \; -- without Variance Data & 0.1899 $\pm$ 0.0233 & 0.4412 $\pm$ 0.0269 & 7.1494 $\pm$ 0.0950 & 81.6464 $\pm$ 1.7613 \\
& \; -- without Residual & 0.1539 $\pm$ 0.0487 & 0.4041 $\pm$ 0.0613 & 7.2739 $\pm$ 0.1937 & 85.3054 $\pm$ 5.0049 \\
& \; -- Covariates Only & 0.1233 $\pm$ 0.0222 & 0.3560 $\pm$ 0.0310 & 7.4076 $\pm$ 0.0450 & 88.3668 $\pm$ 1.8614 \\
\midrule
\textbf{\bm{$G_c$}} 
& SPRN (Proposed) & \textbf{0.3361 $\pm$ 0.0250} & \textbf{0.5833 $\pm$ 0.0195} & 4.2274 $\pm$ 0.1878 & \textbf{29.2638 $\pm$ 2.8185} \\
& \; -- without Variance Data & 0.3358 $\pm$ 0.0237 & 0.5818 $\pm$ 0.0203 & \textbf{4.2278 $\pm$ 0.1955} & 29.2777 $\pm$ 2.7990 \\
& \; -- without Residual & 0.2447 $\pm$ 0.0412 & 0.5373 $\pm$ 0.0251 & 4.5135 $\pm$ 0.1281 & 33.1895 $\pm$ 2.1413 \\
& \; -- Covariates Only & 0.1750 $\pm$ 0.0135 & 0.4237 $\pm$ 0.0163 & 4.6896 $\pm$ 0.1515 & 36.3145 $\pm$ 2.6419 \\
\midrule
\textbf{\bm{$G_t$}} 
& SPRN (Proposed) & 0.3123 $\pm$ 0.0129 & 0.5619 $\pm$ 0.0133 & 5.5462 $\pm$ 0.1121 & 49.3207 $\pm$ 1.3148 \\
& \; -- without Variance Data & \textbf{0.3136 $\pm$ 0.0155} & \textbf{0.5635 $\pm$ 0.0127} & \textbf{5.5417 $\pm$ 0.1002} & \textbf{49.2183 $\pm$ 1.0358} \\
& \; -- without Residual & 0.2521 $\pm$ 0.0286 & 0.5386 $\pm$ 0.0211 & 5.8118 $\pm$ 0.1756 & 53.6651 $\pm$ 2.8399 \\
& \; -- Covariates Only & 0.1914 $\pm$ 0.0236 & 0.4410 $\pm$ 0.0259 & 6.0163 $\pm$ 0.0923 & 57.9765 $\pm$ 1.5624 \\
\bottomrule
\end{tabular*}
\vspace{2mm}

\end{table*}

\begin{table*}[t]
\caption{Sensitivity Analysis of Bootstrapped Feature Selection Thresholds on SPRN Performance. Mean $\pm$ SD across 5 Folds is reported.}
\label{tab:feature_selection}
\centering
\renewcommand{\arraystretch}{1.2}
\begin{tabular*}{\textwidth}{@{\extracolsep{\fill}} l l c c c c}
\toprule
\textbf{Target} & \textbf{Bootstrap Selection Threshold} & \textbf{\bm{$R^2 \uparrow$}} & \textbf{r $\bm{\uparrow}$} & \textbf{MAE $\bm{\downarrow}$} & \textbf{MSE $\bm{\downarrow}$} \\
\midrule
\textbf{\bm{$G_f$}} 
& 30\% Consensus & 0.1822 $\pm$ 0.0178 & 0.4342 $\pm$ 0.0163 & 7.1658 $\pm$ 0.0886 & 82.4343 $\pm$ 1.4780 \\
& 40\% Consensus & 0.1915 $\pm$ 0.0220 & 0.4416 $\pm$ 0.0234 & \textbf{7.1276 $\pm$ 0.0468} & 81.4923 $\pm$ 1.5167 \\
& 50\% Consensus (Proposed) & 0.1906 $\pm$ 0.0175 & 0.4421 $\pm$ 0.0198 & 7.1437 $\pm$ 0.0850 & 81.5946 $\pm$ 1.7941 \\
& 60\% Consensus & \textbf{0.1948 $\pm$ 0.0215} & \textbf{0.4480 $\pm$ 0.0208} & 7.1310 $\pm$ 0.0848 & \textbf{81.1674 $\pm$ 1.9071} \\
& 70\% Consensus & 0.1912 $\pm$ 0.0246 & 0.4423 $\pm$ 0.0290 & 7.1330 $\pm$ 0.1068 & 81.5256 $\pm$ 2.2796 \\
& 80\% Consensus & 0.1879 $\pm$ 0.0253 & 0.4379 $\pm$ 0.0293 & 7.1365 $\pm$ 0.0905 & 81.8505 $\pm$ 1.8591 \\
& 100\% Features (No Selection) & 0.1609 $\pm$ 0.0267 & 0.4094 $\pm$ 0.0261 & 7.2623 $\pm$ 0.1080 & 84.6006 $\pm$ 2.9533 \\
\midrule
\textbf{\bm{$G_c$}} 
& 30\% Consensus & 0.3354 $\pm$ 0.0294 & 0.5815 $\pm$ 0.0237 & \textbf{4.2253 $\pm$ 0.2056} & 29.3040 $\pm$ 3.0252 \\
& 40\% Consensus & 0.3356 $\pm$ 0.0265 & 0.5812 $\pm$ 0.0222 & 4.2374 $\pm$ 0.1981 & 29.2912 $\pm$ 2.9192 \\
& 50\% Consensus (Proposed) & \textbf{0.3361 $\pm$ 0.0250} & \textbf{0.5833 $\pm$ 0.0195} & 4.2274 $\pm$ 0.1878 & \textbf{29.2638 $\pm$ 2.8185} \\
& 60\% Consensus & 0.3342 $\pm$ 0.0234 & 0.5810 $\pm$ 0.0207 & 4.2350 $\pm$ 0.1792 & 29.3306 $\pm$ 2.5758 \\
& 70\% Consensus & 0.3254 $\pm$ 0.0211 & 0.5751 $\pm$ 0.0205 & 4.2539 $\pm$ 0.1714 & 29.7205 $\pm$ 2.6210 \\
& 80\% Consensus & 0.3116 $\pm$ 0.0201 & 0.5639 $\pm$ 0.0177 & 4.2920 $\pm$ 0.1716 & 30.3376 $\pm$ 2.7325 \\
& 100\% Features (No Selection) & 0.3134 $\pm$ 0.0239 & 0.5630 $\pm$ 0.0195 & 4.2992 $\pm$ 0.2202 & 30.2835 $\pm$ 3.0753 \\
\midrule
\textbf{\bm{$G_t$}} 
& 30\% Consensus & \textbf{0.3150 $\pm$ 0.0148} & \textbf{0.5638 $\pm$ 0.0129} & \textbf{5.5432 $\pm$ 0.1246} & \textbf{49.1240 $\pm$ 1.3920} \\
& 40\% Consensus & 0.3137 $\pm$ 0.0200 & 0.5635 $\pm$ 0.0160 & 5.5472 $\pm$ 0.1214 & 49.2090 $\pm$ 1.4459 \\
& 50\% Consensus (Proposed) & 0.3123 $\pm$ 0.0129 & 0.5619 $\pm$ 0.0133 & 5.5462 $\pm$ 0.1121 & 49.3207 $\pm$ 1.3148 \\
& 60\% Consensus & 0.3106 $\pm$ 0.0181 & 0.5612 $\pm$ 0.0142 & 5.5485 $\pm$ 0.0893 & 49.4299 $\pm$ 1.1548 \\
& 70\% Consensus & 0.3123 $\pm$ 0.0207 & 0.5620 $\pm$ 0.0164 & 5.5505 $\pm$ 0.0634 & 49.2901 $\pm$ 0.5093 \\
& 80\% Consensus & 0.3032 $\pm$ 0.0145 & 0.5544 $\pm$ 0.0113 & 5.5727 $\pm$ 0.0840 & 49.9615 $\pm$ 1.0445 \\
& 100\% Features (No Selection) & 0.2927 $\pm$ 0.0095 & 0.5437 $\pm$ 0.0101 & 5.6424 $\pm$ 0.1270 & 50.7370 $\pm$ 1.5289 \\
\bottomrule
\end{tabular*}
\vspace{2mm}

\end{table*}

\begin{table*}[t]
\caption{Comparison of Predictive Performance Across Methodological Families. Mean $\pm$ SD across 5 Folds is reported.}
\label{tab:method_comparison}
\centering
\renewcommand{\arraystretch}{1.2}
\begin{tabular*}{\textwidth}{@{\extracolsep{\fill}} l l c c c c}
\toprule
\textbf{Target} & \textbf{Method} & \textbf{\bm{$R^2 \uparrow$}} & \textbf{r $\bm{\uparrow}$} & \textbf{MAE $\bm{\downarrow}$} & \textbf{MSE $\bm{\downarrow}$} \\
\midrule
\textbf{\bm{$G_f$}} 
& Ridge Regression & 0.1028 $\pm$ 0.0279 & 0.3684 $\pm$ 0.0243 & 7.5259 $\pm$ 0.1298 & 90.4258 $\pm$ 2.1422 \\
& Lasso Regression & -0.0087 $\pm$ 0.0055 & 0.0000 $\pm$ 0.0000 & 7.9626 $\pm$ 0.1329 & 101.7041 $\pm$ 1.8196 \\
& ElasticNet & -0.1361 $\pm$ 0.0386 & 0.3061 $\pm$ 0.0255 & 8.4998 $\pm$ 0.1611 & 114.5058 $\pm$ 3.1759 \\
& Random Forest & 0.1256 $\pm$ 0.0136 & 0.3688 $\pm$ 0.0187 & 7.4148 $\pm$ 0.0821 & 88.1493 $\pm$ 1.6804 \\
& Gradient Boosting & 0.1393 $\pm$ 0.0098 & 0.3811 $\pm$ 0.0105 & 7.3559 $\pm$ 0.0760 & 86.7708 $\pm$ 1.0400 \\
& Standard GCN & 0.1289 $\pm$ 0.0192 & 0.3625 $\pm$ 0.0258 & 7.3879 $\pm$ 0.0597 & 87.8054 $\pm$ 1.6192 \\
& \textbf{SPRN (Proposed)} & \textbf{0.1906 $\pm$ 0.0175} & \textbf{0.4421 $\pm$ 0.0198} & \textbf{7.1437 $\pm$ 0.0850} & \textbf{81.5946 $\pm$ 1.7941} \\
\midrule
\textbf{\bm{$G_c$}} 
& Ridge Regression & 0.2670 $\pm$ 0.0332 & 0.5237 $\pm$ 0.0282 & 4.4762 $\pm$ 0.2304 & 32.3274 $\pm$ 3.3930 \\
& Lasso Regression & -0.0033 $\pm$ 0.0059 & 0.0000 $\pm$ 0.0000 & 5.1703 $\pm$ 0.2007 & 44.1837 $\pm$ 3.4029 \\
& ElasticNet & 0.0992 $\pm$ 0.0458 & 0.4663 $\pm$ 0.0296 & 4.9606 $\pm$ 0.2400 & 39.6689 $\pm$ 3.6908 \\
& Random Forest & 0.1864 $\pm$ 0.0159 & 0.4599 $\pm$ 0.0202 & 4.6410 $\pm$ 0.1744 & 35.8478 $\pm$ 3.0561 \\
& Gradient Boosting & 0.2417 $\pm$ 0.0225 & 0.4965 $\pm$ 0.0216 & 4.4991 $\pm$ 0.1813 & 33.4149 $\pm$ 3.0206 \\
& Standard GCN & 0.1743 $\pm$ 0.0134 & 0.4230 $\pm$ 0.0187 & 4.6990 $\pm$ 0.1528 & 36.3311 $\pm$ 2.4469 \\
& \textbf{SPRN (Proposed)} & \textbf{0.3361 $\pm$ 0.0250} & \textbf{0.5833 $\pm$ 0.0195} & \textbf{4.2274 $\pm$ 0.1878} & \textbf{29.2638 $\pm$ 2.8185} \\
\midrule
\textbf{\bm{$G_t$}} 
& Ridge Regression & 0.2376 $\pm$ 0.0197 & 0.4998 $\pm$ 0.0186 & 5.8770 $\pm$ 0.1355 & 54.6926 $\pm$ 2.1475 \\
& Lasso Regression & -0.0036 $\pm$ 0.0016 & 0.0000 $\pm$ 0.0000 & 6.6737 $\pm$ 0.1720 & 72.0015 $\pm$ 2.4863 \\
& ElasticNet & 0.0700 $\pm$ 0.0399 & 0.4434 $\pm$ 0.0295 & 6.4856 $\pm$ 0.1647 & 66.6987 $\pm$ 3.2993 \\
& Random Forest & 0.2003 $\pm$ 0.0178 & 0.4720 $\pm$ 0.0232 & 5.9651 $\pm$ 0.1492 & 57.3574 $\pm$ 1.8635 \\
& Gradient Boosting & 0.2437 $\pm$ 0.0169 & 0.4970 $\pm$ 0.0147 & 5.8121 $\pm$ 0.1436 & 54.2444 $\pm$ 1.8389 \\
& Standard GCN & 0.1996 $\pm$ 0.0225 & 0.4498 $\pm$ 0.0248 & 5.9894 $\pm$ 0.1042 & 57.3911 $\pm$ 1.5639 \\
& \textbf{SPRN (Proposed)} & \textbf{0.3123 $\pm$ 0.0129} & \textbf{0.5619 $\pm$ 0.0133} & \textbf{5.5462 $\pm$ 0.1121} & \textbf{49.3207 $\pm$ 1.3148} \\
\bottomrule
\end{tabular*}
\vspace{2mm}

\parbox{1\textwidth}{\small \textit{Note:} All methods use the identical 50\% consensus mask. Bold indicates best per target. \(\uparrow/\downarrow\): higher/lower is better.}
\end{table*}

\begin{table*}[t]
\caption{State-of-the-Art Comparison of Intelligence Prediction using rs-fMRI on the ABCD Dataset.}
\label{tab:sota_comparison}
\centering
\renewcommand{\arraystretch}{1.2}
\begin{tabular*}{\textwidth}{@{\extracolsep{\fill}} l l l c c c}
\toprule
\textbf{Study} & \textbf{Model Architecture} & \textbf{Modality} & \textbf{\bm{$G_f (r)$}} & \textbf{\bm{$G_c (r)$}} & \textbf{\bm{$G_t (r)$}} \\
\midrule
Huang et al. (2022) \cite{Huang_2022} & ST-DAG-Att & rs-fMRI & 0.288 & - & - \\
Li et al. (2023) \cite{Li_2023} & Bi-LSTM (Dynamic FC) & rs-fMRI (Rest Only) & 0.408 & 0.536 & 0.529 \\
Xia et al. (2023) \cite{Xia_2023} & ML-Joint-Att & rs-fMRI & 0.298 & 0.391 & - \\
Thapaliya et al. (2025) \cite{fc_model3} & BrainRGIN & rs-fMRI & 0.230 & 0.300 & 0.310 \\
\midrule
\textbf{This Study} & \textbf{SPRN (Proposed)} & \textbf{rs-fMRI} & \textbf{0.442} & \textbf{0.583} & \textbf{0.562} \\
\bottomrule
\end{tabular*}
\vspace{2mm}

\end{table*}

\section{Results and Discussion}
The primary objective of this study was to evaluate the efficacy of the SPRN in predicting human intelligence from resting-state functional connectomes. The comparative predictive performance of our complete ablation suite across $G_f$, $G_c$, and $G_t$ is detailed in Tables \ref{tab:sprn_ablation}, \ref{tab:feature_selection}, and \ref{tab:method_comparison}.

\subsection{Predictive Performance vs. Covariates}
To verify that the learned representations captured genuine neurobiological information rather than demographic or acquisition-related effects, we constructed a strict ``Covariates-Only'' baseline in which all brain-derived features were replaced with Gaussian noise. This constrained the model to predict intelligence using only the covariate set \(\mathbf{c}_i\).

Across all three cognitive domains, the SPRN consistently outperformed the demographic baseline, demonstrating that resting-state functional connectomes contain unique predictive variance beyond nuisance covariates alone. The strongest performance was observed for \(G_c\), where the SPRN achieved \(R^2 = 0.3361\;(r = 0.5833)\), substantially exceeding the baseline (\(R^2 = 0.1750,\; r = 0.4237\)). This suggests that crystallized abilities are strongly reflected in stable, large-scale functional connectivity patterns measurable during rest.

Similarly, the SPRN achieved robust performance for \(G_t\) with \(R^2 = 0.3123\;(r = 0.5619)\), outperforming the baseline (\(R^2 = 0.1914,\; r = 0.4410\)). \(G_f\) remained the most challenging target, likely due to its dependence on dynamic reasoning processes that are less directly captured by static rs-fMRI connectivity. Nevertheless, the model still achieved meaningful predictive performance (\(R^2 = 0.1906,\; r = 0.4421\)) and exceeded the baseline (\(R^2 = 0.1233,\; r = 0.3560\)), indicating that rs-fMRI features provide stable predictive information even for dynamic cognitive abilities.

\subsection{Ablation Analyses}
\subsubsection{The Insufficiency of Traditional and Graph Models}
To benchmark the SPRN against established approaches, we evaluated several standard ML models using the identical 50\% consensus feature mask (Table \ref{tab:method_comparison}). Linear models struggled substantially within the high-dimensional connectomic feature space. Lasso Regression completely failed, producing negative $R^2$ values across all targets, while ElasticNet also yielded negative performance for $G_f$ and $G_t$. These findings suggest that strict linear sparsity assumptions are poorly suited to the dense and highly correlated structure of sFC data.

Non-linear ensemble methods performed more competitively, with Gradient Boosting achieving \(R^2 = 0.2417\) for \(G_c\) and \(R^2 = 0.2437\) for \(G_t\). Although these models exceeded the covariates-only baseline, they remained clearly below the DL approaches, indicating that the predictive structure of the connectome is strongly non-linear and benefits from hierarchical feature learning.

We further evaluated a standard GCN to assess whether graph-based message passing improves prediction performance. The GCN consistently underperformed, achieving only \(R^2 = 0.1743\) for \(G_c\), which was lower than even a Random Forest trained on the same features. This likely reflects the over-smoothing problem commonly observed in deep GNNs on dense brain graphs \cite{oversmoothing1, oversmoothing2}, where repeated message passing progressively reduces fine-grained regional distinctions. In contrast, the SPRN preserves edge-specific variance by treating connectivity edges as independent features within a residual fully connected architecture.

\subsubsection{Robustness of Bootstrapped Dimensionality Reduction}
To assess the stability of the bootstrapped feature selection procedure, we evaluated the SPRN across multiple consensus thresholds ranging from 30\% to 100\% (no feature selection) (Table \ref{tab:feature_selection}). Using the full unreduced connectome consistently degraded performance across all cognitive targets. For example, \(R^2\) for \(G_c\) decreased from 0.3361 (50\% threshold) to 0.3134 without selection, while \(G_t\) declined from 0.3123 to 0.2927. These findings indicate that removing noisy and redundant connections is essential for isolating meaningful neurocognitive signals from high-dimensional connectivity data.

Although minor target-specific improvements emerged at alternative thresholds (e.g., 60\% for \(G_f\) and 30\% for \(G_t\)), performance remained consistently stable within the 40--60\% range. We therefore adopted the 50\% majority consensus threshold as a principled, target-agnostic setting across all experiments. This choice avoids over-optimization to individual cognitive domains while demonstrating that the learned feature masks remain robust across different threshold selections.

\subsubsection{Topology Dominates Over Local Amplitude}
Ablation of the node-level temporal variance features (\(\mathbf{x}_{\text{var}}\)) revealed that local BOLD amplitude contributed minimally to prediction performance (Table \ref{tab:sprn_ablation}). For \(G_f\), removing variance features produced nearly identical performance (\(R^2 = 0.1899\) vs.\ \(0.1906\)), while \(G_c\) also remained unchanged (\(R^2 = 0.3358\) vs.\ \(0.3361\)). Interestingly, performance for \(G_t\) slightly improved after excluding variance features (\(R^2 = 0.3136\) vs.\ \(0.3123\)), suggesting that nodal amplitude may introduce mild noise into the prediction of global intelligence.

These findings suggest that intelligence is more strongly encoded in the large-scale topology and synchrony of functional networks than in isolated regional activity. While local BOLD variance has been associated with task-related activation \cite{Kaboodvand2020}, its contribution to resting-state intelligence prediction appears limited when comprehensive sFC information is available. Overall, the results support network neuroscience perspectives that emphasize distributed functional interactions as the primary substrate of cognition.

\subsubsection{The Necessity of Residual Projections}
Removing the learned projection residuals produced the largest performance degradation across all architectural ablations (Table \ref{tab:sprn_ablation}). Without the projected skip connections, the SPRN effectively reduced to a standard MLP. Performance for \(G_c\) dropped sharply from \(R^2 = 0.3361\) to \(0.2447\), while \(G_f\) declined from \(0.1906\) to \(0.1539\), and \(G_t\) decreased from \(0.3123\) to \(0.2521\). These results demonstrate that residual projections are critical for preserving predictive information through the aggressive 192-to-24 neuron bottleneck.

The projection matrices (\({W}_{\text{proj}, l}\)) provide two key benefits. First, they improve gradient flow and stabilize optimization during deep compression. Second, they encourage each layer to learn only the residual information beyond a simple linear transformation of the previous representation. This residual learning mechanism helps preserve fine-grained connectomic structure throughout the compression hierarchy, explaining the substantial performance loss observed when the projections were removed.

\subsection{Comparison with State-of-the-Art Architectures}
To contextualize the predictive capacity of the proposed SPRN, we compared its performance against recent state-of-the-art DL models evaluated on the ABCD dataset (Table \ref{tab:sota_comparison}). Earlier graph-based approaches reported relatively modest performance for \(G_f\) (\(r = 0.288\) \cite{Huang_2022} and \(r = 0.298\) \cite{Xia_2023}), reflecting the difficulty of modeling fluid reasoning from static connectivity alone. More recent graph isomorphism network approaches similarly achieved moderate correlations across cognitive targets (\(r = 0.230\)--\(0.310\)) \cite{fc_model3}. To address the limitations of static graph representations, dynamic functional connectivity models based on bidirectional long short-term memory networks achieved stronger performance (\(r = 0.408\) for \(G_f\), \(r = 0.529\) for \(G_t\), and \(r = 0.536\) for \(G_c\)) by explicitly modeling temporal fluctuations in connectivity \cite{Li_2023}.

Despite relying exclusively on sFC, the SPRN achieved superior performance across all three cognitive domains, reaching \(r = 0.442\) for \(G_f\), \(0.583\) for \(G_c\), and \(0.562\) for \(G_t\). Unlike dynamic sequence-based architectures, which require substantial computational overhead to process windowed connectivity trajectories, the SPRN operates on a compact, aggressively pruned static feature space derived via bootstrapped Ridge consensus masking. Combined with the proposed wide-to-narrow residual architecture, this design demonstrates that careful feature selection and residual representation learning can recover highly informative cognitive signatures from static connectomic structure alone.

Direct statistical comparisons across studies remain difficult because prior studies used different ABCD cohort subsets and preprocessing pipelines. Nevertheless, the SPRN's consistent performance across all three intelligence domains within a strictly controlled nested cross-validation framework supports the robustness and effectiveness of the proposed architecture.

\begin{figure*} [!t]
	\centering
	\includegraphics[width=\textwidth]{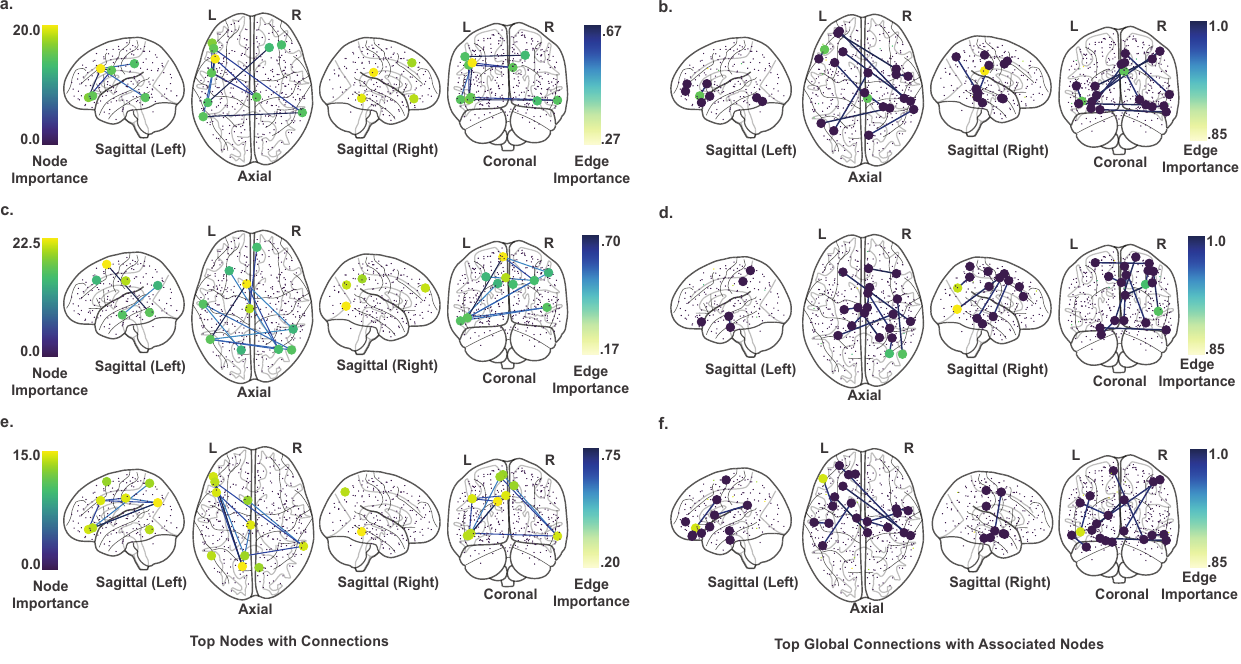}
	\caption{XAI Consensus visualization of neuroanatomical drivers for intelligence prediction. Left panels (a, c, e): Top 10 hubs (nodes) and their strongest inter-connections. Right panels (b, d, f): Top 10 global edges across the whole brain. Color intensity reflects attribution magnitude. Panels correspond to $G_f$ (a, b), $G_c$ (c, d), and $G_t$ (e, f).}
	\label{fig:xai}
\end{figure*}

\begin{table*}[t]
\caption{Multi-Algorithm Convergence Analysis of XAI Feature Attributions.}
\label{tab:xai_convergence}
\centering
\renewcommand{\arraystretch}{1.2}
\begin{tabular*}{\textwidth}{@{\extracolsep{\fill}} l l l l l l}
\toprule
\textbf{Cognitive Target} & \textbf{Explainer Pair} & \textbf{Spearman $\bm{\rho}$} & \textbf{Kendall $\bm{\tau}$} & \textbf{Pearson \textit{r}} & \textbf{Jaccard (Top 100)} \\
\midrule
\textbf{$G_f$} 
& IG vs. GradientSHAP & 0.9998 & 0.9874 & 0.9999 & 0.9802 \\
& IG vs. Occlusion    & 0.9995 & 0.9857 & 0.9997 & 0.9802 \\
& GradientSHAP vs. Occlusion & 0.9993 & 0.9809 & 0.9996 & 0.9608 \\
\midrule
\textbf{$G_c$} 
& IG vs. GradientSHAP & 0.9998 & 0.9883 & 0.9999 & 0.9802 \\
& IG vs. Occlusion    & 0.9992 & 0.9829 & 0.9996 & 0.9802 \\
& GradientSHAP vs. Occlusion & 0.9990 & 0.9793 & 0.9995 & 0.9802 \\
\midrule
\textbf{$G_t$} 
& IG vs. GradientSHAP & 0.9997 & 0.9869 & 0.9998 & 0.9608 \\
& IG vs. Occlusion    & 0.9992 & 0.9831 & 0.9996 & 0.9417 \\
& GradientSHAP vs. Occlusion & 0.9990 & 0.9780 & 0.9995 & 0.9608 \\
\bottomrule
\end{tabular*}
\vspace{2mm}

\end{table*}

\subsection{Explainable AI and Connectomic Biomarkers}

\subsubsection{Multi-Algorithm Convergence}

A major concern in DL-based neuroimaging is the reliability of feature attribution methods, as different explainers can often produce inconsistent interpretations. However, as detailed in Table \ref{tab:xai_convergence}, the attribution maps demonstrated exceptionally strong convergence across all cognitive domains. Attribution magnitudes were nearly identical across methods (Pearson \(r \ge 0.9995\)), while feature rankings remained highly stable, with Spearman \(\rho \ge 0.9990\) and Kendall \(\tau \ge 0.9780\) across all explainer pairs. Furthermore, the overlap among the Top-100 most important pathways remained extremely high (\(\text{Jaccard} \ge 0.9417\)), confirming that all attribution methods consistently identified the same core set of predictive connections.

This exceptional convergence ($\rho > 0.999$) across distinct gradient, attribution, and perturbation paradigms is a direct mathematical consequence of our 50\% bootstrapped feature masking pipeline. By aggressively pruning topological noise to isolate the top $\sim$3.5\% of predictive edges, the optimization landscape is highly constrained. Devoid of spurious features, the independent XAI algorithms were structurally forced through identical topological bottlenecks, yielding near-identical importance configurations. This convergence not only validates the feature selection pipeline but also substantially strengthens the reliability of the neurobiological interpretations that follow.

\subsubsection{Topological Routing: Hubs vs. Highways}
To systematically interpret the predictive topography of intelligence, we adopted a dual-layered analytical framework that distinguishes between Node Importance and Global Edge Importance. Node Importance quantifies centralized processing hubs and is calculated as the sum of all normalized edge attributions incident to a given brain region, augmented by that region's intrinsic temporal variance attribution score. Global Edge Importance, by contrast, captures distinct, high-bandwidth connectomic highways, which are the individual functional connections carrying the highest predictive weight across the entire brain graph, irrespective of whether their terminal nodes rank among the top hubs.

This dual-layered architecture is visualized in Fig. \ref{fig:xai}, where the left panels display the highly localized primary hubs and their core interconnections, while the right panels reveal the distinct, long-range global routing strategies that support each cognitive domain. Across all three intelligence facets, a consistent and neurobiologically revealing pattern emerged: while primary processing hubs are densely localized within specific canonical networks, the absolute strongest global connections frequently bypass these primary nodes, instead utilizing distinct relay regions to orchestrate whole-brain communication. This hub-versus-highway dissociation suggests that intelligence does not simply reflect the activity of a fixed set of important regions but rather depends on a flexible routing architecture in which centralized hubs and decentralized long-range pathways serve complementary computational roles.

\subsubsection{Domain-Specific Mappings}

The neuroanatomical signatures of \(G_f\) strongly align with the Parieto-Frontal Integration Theory (P-FIT) \cite{pfit}. The most influential hubs were concentrated within the Left Fronto-Parietal, Left Ventral Attention, and Left Dorsal Attention networks, with strong intra-network and bilateral Fronto-Parietal coupling. Notably, the highest-weighted global edge (consensus score: 1.0) was a direct interhemispheric connection between the left and Right Fronto-Parietal regions. Although these terminal nodes were not ranked among the strongest hubs, their dominance as a global pathway suggests that fluid reasoning depends on efficient, decentralized bilateral communication that supports flexible problem-solving.

For \(G_c\), the predictive topology was strongly left-lateralized and dominated by Ventral Attention and Default Mode hubs, consistent with distributed semantic retrieval and language-related processing. However, the strongest global pathways selectively recruited right-hemispheric and subcortical relay regions. The top-ranked edge connected the Right Salience and Right Fronto-Parietal networks, while other prominent pathways involved subcortical relays such as the Right Sensorimotor to Left Ventral Diencephalon connection. These findings suggest that crystallized intelligence relies not only on localized cortical knowledge representations but also on specialized relay pathways that coordinate access to distributed semantic information.

As a composite measure of general intelligence, \(G_t\) exhibited a broader integrative architecture spanning Left Cingulo-Parietal, Left Default Mode, and bilateral Fronto-Parietal systems. A notable emergent feature was the dominance of the Left Cingulo-Parietal hub, which was the only highly ranked node across all domains to retain a meaningful positive variance contribution. While the SPRN largely bypassed local BOLD amplitude information elsewhere, this result suggests that regional amplitude modulation within this task-control hub may remain relevant for higher-order cognitive integration. At the global level, \(G_t\) was strongly driven by cross-network communication between the Left Salience and Right Cingulo-Parietal systems. In addition, several top-ranked pathways uniquely implicated cortico-striatal circuits involving the Left Caudate nucleus, highlighting a potential role for basal ganglia coordination in integrating distributed cognitive processes into unified intellectual function.

\subsection{Clinical and Scientific Translation}
From a translational perspective, the consensus XAI framework provides biologically grounded insights regarding the functional organization of intelligence. A key insight was the minimal contribution of node-level temporal variance once sFC information was available. This is quantitatively supported by our ablation studies, which showed negligible performance drops when variance features were removed. This suggests that topology-based biomarkers may be more useful than amplitude-based measures for future cognitive and clinical modeling \cite{woo2017building}. Furthermore, the identified connectomic signatures align with established clinical observations. The strong left-lateralization of Default Mode and Ventral Attention hubs for predicting \(G_c\) is highly consistent with language and semantic impairments observed following left-hemispheric disruptions, such as stroke or temporal lobe epilepsy \cite{Duncan_Pradeep2025}. Consequently, these network-specific signatures could serve as neuroimaging-based markers for neurodegenerative decline, where early hub degradation may inform prognosis and guide rehabilitation planning.

Additionally, the systematic dissociation between localized computational hubs and long-range global routing pathways across all cognitive domains supports a network-based view of intelligence. This indicates that efficient large-scale communication may be as critical to general cognitive function as localized cortical processing \cite{popp2024structural}. While all associations reported in this study are strictly predictive rather than causal, these findings demonstrate how explainable deep learning can extend traditional predictive neuroimaging by translating high-dimensional connectivity patterns into specific, testable network-level models regarding human cognition. Clinically, the ability to isolate these distinct hub-and-relay pathways offers a mechanism to identify targeted intervention points for neurocognitive disorders.

\subsection{Limitations and Future Directions}
Despite the strong predictive performance of the SPRN framework, several limitations remain. First, the model relies exclusively on static functional connectivity (sFC) averaged over the full scan duration. While this design minimizes computational complexity compared to dynamic sequence models, it does not capture moment-to-moment neural state transitions. Future work should explore lightweight attention mechanisms to incorporate temporal connectivity dynamics without the steep overhead of recurrent or transformer architectures. Furthermore, recent brain representation learning paradigms indicate that adaptive token selection and targeted pretraining strategies can isolate subtle brain-behavior associations in high-dimensional neuroimaging data \cite{Jiang2024}. Similarly, hypercomplex graph learning offers a potential framework for modeling multi-relational interactions across complementary brain networks while preserving network-specific structures \cite{Yang2025}.

Second, this study's cross-sectional design limits causal interpretation, as the identified connectomic patterns represent predictive associations rather than confirmed mechanisms driving intelligence. Although the ABCD dataset provides longitudinal waves, this study explicitly utilized baseline cross-sectional data to rigorously validate the novel SPRN architecture and XAI consensus pipeline without introducing the confounding variance of temporal developmental shifts. Longitudinal tracking is now a critical next step to determine whether these hub-and-highway architectures actively influence cognitive development or merely reflect stable individual differences, particularly since the ABCD cohort consists primarily of US-based children and adolescents. Additionally, while our explainability framework demonstrated near-perfect statistical convergence across Integrated Gradients, GradientSHAP, and Occlusion, post-hoc XAI scores remain descriptive measures of model dependence rather than direct indicators of neurobiological necessity. Future studies combining predictive modeling with longitudinal monitoring or non-invasive brain stimulation techniques—such as TMS or tDCS—are required to validate the causal relevance of the identified topologies.

\section{Conclusion}
We presented SPRN, an explainable DL framework for predicting \(G_f\), \(G_c\), and \(G_t\) from resting-state functional connectomes. By combining bootstrapped consensus sparsification with learned residual projections, SPRN mitigates the instability and over-smoothing commonly observed in standard deep graph models \cite{glorot2010understanding, peng2024beyond}, achieving state-of-the-art predictive performance relative to existing connectomic benchmarks across all cognitive domains (mean \(r = 0.529\)).

Beyond prediction, the proposed multi-algorithm XAI framework demonstrated near-perfect convergence across IG, GradientSHAP, and Occlusion (\(\rho > 0.999\)), indicating that the identified neurobiological patterns were highly stable rather than algorithm-specific artifacts. The resulting attribution maps revealed a dual organizational structure of intelligence: while dominant computational hubs were localized within canonical networks (e.g., Fronto-Parietal systems for \(G_f\) and left-lateralized DMN regions for \(G_c\)), the strongest predictive pathways frequently relied on distributed long-range relay connections. In particular, \(G_c\) and \(G_t\) prominently recruited right-lateralized and subcortical cortico-striatal pathways, highlighting the importance of large-scale communication architecture in cognitive organization.

Although these findings remain predictive rather than causal, the strict control of demographic and acquisition-related confounds strengthens the interpretation that the identified patterns reflect behaviorally relevant neural organization. Ultimately, SPRN provides a robust and interpretable framework for translating high-dimensional rs-fMRI connectivity into biologically meaningful insights about individual differences in human intelligence. This transparent mapping establishes a normative network baseline, offering a powerful computational foundation for detecting atypical neurodevelopment and guiding personalized, network-targeted therapies.

\section*{Acknowledgment}
This research was supported by the Science and Technology Fellowship Trust, Government of the People's Republic of Bangladesh, and the Commonwealth through an Australian Government Research Training Program Scholarship. Data used in the preparation of this article were obtained from the Adolescent Brain Cognitive Development (ABCD) Study, held in the NIH Brain Development Cohorts Data Sharing Platform. In addition, this work was supported by resources provided by The University of Queensland Research Computing Center’s Bunya supercomputer.

\section*{Code \& Data Availability}
Codes associated with this study will be made publicly available after peer review. The ABCD Study data utilized in this research are available through the National Institute of Mental Health Data Archive (NDA) to approved researchers. Access requires an executed Data Use Certification agreement (https://nda.nih.gov/abcd)

\printbibliography

\end{document}